\documentclass[aps,prd,twocolumn,nofootinbib,superscriptaddress,secnumarabic]{revtex4-2}

\usepackage{amsmath,amssymb,graphicx,xcolor}
\usepackage[normalem]{ulem}
\usepackage[colorlinks=true,urlcolor=blue,anchorcolor=blue,citecolor=blue,filecolor=blue,linkcolor=blue,menucolor=blue,linktocpage=true,pdfproducer=medialab,pdfa=true]{hyperref}
\usepackage{orcidlink}

\begin{document}
\title{Primordial Black Hole interpretation of the subsolar merger event S251112cm}

\author{Md Riajul Haque \orcidlink{0000-0002-6618-4899}}
\email{riaj1994@sjtu.edu.cn}
\affiliation{Tsung-Dao Lee Institute \& School of Physics and Astronomy, Shanghai Jiao Tong University, Shanghai 201210, China}

\author{Fabio Iocco \orcidlink{0000-0002-4237-0005}}
\email{fabio.iocco.unina@gmail.com}
\affiliation{Universit\`a di Napoli ``Federico II'' \& INFN Napoli, via Cintia, 80126 Napoli, Italy}

\author{Luca Visinelli \orcidlink{0000-0001-7958-8940}}
\email{lvisinelli@unisa.it}
\affiliation{Dipartimento di Fisica ``E.R.\ Caianiello'', Universit\`a degli Studi di Salerno,\\ Via Giovanni Paolo II, 132 - 84084 Fisciano (SA), Italy}
\affiliation{Istituto Nazionale di Fisica Nucleare - Gruppo Collegato di Salerno - Sezione di Napoli,\\ Via Giovanni Paolo II, 132 - 84084 Fisciano (SA), Italy}

\date{\today}

\begin{abstract}
The LIGO-Virgo-KAGRA (LVK) candidate event S251112cm suggests that at least one component has a subsolar mass. Since such objects cannot be produced through standard stellar evolution, this observation provides a possible indication of nonstandard formation channels. Primordial black holes (PBHs), formed from the collapse of primordial density fluctuations in the early Universe, are a well-motivated candidate. We investigate the interpretation of S251112cm as the merger of two PBHs in the mass range $0.20$--$1\,M_\odot$, within the parameter space directly covered by the O4a subsolar search. Combining analytic estimates of the PBH merger rate with current observational constraints on their abundance and the sensitivity of LVK searches, we compute the probability of observing such an event. We consider both a monochromatic PBH mass function and extended lognormal distributions. Within the relaxed constraint scenario, the detection probability can approach unity toward the upper end of the mass range. A lognormal mass distribution of moderate width can enhance the probability relative to the monochromatic case, whereas a broader distribution can instead suppress it. Our results show that a PBH interpretation of S251112cm is compatible with current observational bounds even for a subdominant PBH abundance. Owing to the dependence of our results on astrophysical uncertainties, such as those affecting the constraints on the abundance of PBHs, they cannot be regarded as conclusive with respect to the nature of the detected event. At the same time, our analysis highlights the potential of subsolar gravitational wave events as a probe of PBHs and their contribution to dark matter.
\end{abstract}

\maketitle

\section{Introduction}

The LIGO--Virgo--KAGRA (LVK) collaboration has recently reported the gravitational wave (GW) candidate event S251112cm, observed during the O4 run in November 2025~\cite{2025GCN.42650....1L, 2025GCN.42690....1L}. Preliminary parameter estimation indicates a chirp mass in the range $\sim 0.1$--$0.9\,M_\odot$~\cite{Vieira:2026eof}, implying that at least one component has a high probability of being subsolar. The event is located at a luminosity distance of $\sim 90\,\mathrm{Mpc}$ and has a false alarm rate of $\sim 1$ per few years, while no confirmed electromagnetic counterpart has been established. However, Ref.~\cite{Hall:2026gov} reports a spatially and temporally coincident Type IIb supernova, SN~2025adtq, providing suggestive but inconclusive evidence for an association with S251112cm. An independent near-infrared follow-up campaign found no convincing transient counterpart~\cite{Paek:2026kat}. The possible observation of a compact object in the subsolar mass range is of particular interest, as such objects cannot be formed through standard stellar evolution channels~\cite{Carr:2020xqk}. Primordial black holes (PBHs) provide a natural explanation for compact objects below the Chandrasekhar limit. They can form in the early Universe from the collapse of large density fluctuations, independently of stellar processes~\cite{Zeldovich:1967lct,Hawking:1971ei,Carr:1974nx,Carr:1975qj}. In this scenario, the PBH mass is set by the horizon mass at formation, establishing a connection between the formation epoch and the PBH mass scale. Subsolar PBHs can arise from perturbations collapsing around the QCD epoch, where the equation of state enhances collapse efficiency~\cite{Jedamzik:1996mr,Khlopov:2008qy,Byrnes:2018clq,Musco:2020jjb}, producing masses $\lesssim 1\,M_\odot$~\cite{Magaraggia:2026jhk}.\footnote{See Refs.~\cite{Dasgupta:2020mqg, Chen:2025uwd} for alternative mechanisms.} Alternative mechanisms involving substantial PBH mass growth in the early Universe can also produce objects in the solar mass range~\cite{Vitale:2026ric}.

PBHs are subject to Hawking evaporation~\cite{Hawking:1974rv}, but those with masses $M \gtrsim 10^{15}\,\mathrm{g}$ have lifetimes exceeding the age of the Universe and can survive until today, behaving effectively as a dark matter component. Their abundance is constrained by a wide range of astrophysical and cosmological probes, including microlensing, dynamical effects, accretion signatures, and cosmic microwave background measurements; see, e.g., Refs.~\cite{Carr:2009jm,Belotsky:2014kca,Carr:2020gox,Green:2020jor,Escriva:2022duf} for reviews. These constraints are strongly mass-dependent and are typically expressed in terms of the fraction $f_{\rm PBH}$ of dark matter in PBHs. In the mass range $m \sim 0.1$--$1\,M_\odot$, the most stringent bounds arise from microlensing surveys toward the Large Magellanic Cloud~\cite{MACHO:2000qbb,EROS-2:2006ryy,Mroz:2024mse}, although their exact strength depends on the modeling of the Galactic halo and disk. Despite these constraints, a subdominant PBH population remains allowed and can give rise to observable merger rates.

A key theoretical feature of PBHs is that their mass spectrum is not tied to stellar evolution and can span a wide range of masses depending on the primordial power spectrum. This allows for the existence of compact objects in otherwise inaccessible mass windows, including the subsolar region probed by LVK. The detection of a subsolar merger would constitute a qualitatively new probe of early Universe physics and the origin of dark matter. Previous studies have explored the potential of GW observations to probe PBHs through their merger rates and mass distributions~\cite{Nakamura:1997sm,Bird:2016dcv,Clesse:2016vqa, Sasaki:2016jop,Ali-Haimoud:2017rtz,Raidal:2018bbj,Belotsky:2018wph, Mukherjee:2021ags,Boybeyi:2024aax,Boccia:2024nly,Yuan:2024yyo, Domenech:2026nun}.

Importantly, extending the PBH mass function beyond the monochromatic approximation modifies both the observational constraints on the PBH abundance and the predicted merger rate~\cite{Kuhnel:2017pwq, Carr:2017jsz, Bellomo:2017zsr, Raidal:2018bbj}. Consequently, the impact of an extended mass distribution on the detectable merger probability must be assessed consistently, accounting for both the mass-dependent abundance bounds and binary-suppression effects. More generally, the PBH mass function encodes detailed information about the early Universe dynamics responsible for primordial fluctuations. Extended mass distributions, such as lognormal spectra arising from enhanced small-scale perturbations, can simultaneously account for observables across widely separated mass scales~\cite{Yuan:2023bvh,Cai:2023ptf}. In particular, PBHs with masses up to $\sim 10^3\,M_\odot$ have been proposed as seeds for the rapid formation of massive high-redshift galaxies observed by the James Webb Space Telescope (JWST), favoring broad mass functions over monochromatic ones~\cite{Yuan:2023bvh}. Conversely, the shape of the PBH mass function itself can serve as a probe of the underlying early Universe scenario, allowing one to discriminate between different mechanisms such as small-scale-enhanced inflation or matter-bounce cosmologies~\cite{Cai:2023ptf}. In this context, subsolar mergers offer a particularly promising observational test, as they are largely free from astrophysical backgrounds.

In this {\it Letter}, we investigate whether the S251112cm event can be consistently interpreted as the merger of subsolar mass PBHs. We compute the expected merger rate as a function of $f_{\rm PBH}$ and combine it with the sensitivity of LVK searches to evaluate the probability of observing such an event. We consider both monochromatic and extended lognormal PBH mass functions, incorporating the corresponding merger rate suppression and recast microlensing constraints. Assuming equal mass binaries, we use the O4a detector sensitivity to evaluate the detection probability over the mass interval $0.20\leq m/M_\odot\leq 1$ and investigate its dependence on the characteristic mass, distribution width, and PBH abundance. This allows us to assess the viability of a primordial origin and to quantify the implications for the PBH contribution to dark matter.

\section{Methods}

A compact binary with component masses $m_1$ and $m_2$ is characterized by the chirp mass
\begin{equation}
    \mathcal{M}=\frac{(m_1m_2)^{3/5}}{(m_1+m_2)^{1/5}}\,,
\end{equation}
which governs the phase evolution of the waveform and is the best-constrained mass parameter in GW observations. Introducing the mass ratio $q=m_2/m_1\leq1$, the component masses are
\begin{equation}
    m_1=\mathcal{M}q^{-3/5}(1+q)^{1/5}, \qquad m_2=q m_1\,.
\end{equation}
Throughout our analysis, we assume equal mass binaries, $m_1=m_2=m$, for which $\mathcal{M}=2^{-1/5}m$. The preliminary chirp-mass estimate associated with the candidate event S251112cm is~\cite{Vieira:2026eof}
\begin{equation}
    \mathcal{M}\sim(0.1\text{--}0.9)\,M_\odot\,,
\end{equation}
which lies below the typical stellar-mass binary black hole regime, suggesting that at least one component is subsolar. The component masses depend on the chirp mass and mass ratio, making a PBH interpretation a natural possibility. The differential merger rate of PBH binaries at cosmic time $t$ is given by~\cite{Raidal:2018bbj,Mukherjee:2021ags,Boybeyi:2024aax}
\begin{align}
    \frac{\mathrm{d}R}{\mathrm{d}m_1\,\mathrm{d}m_2} &\approx 1.6\times10^6\,\mathrm{Gpc}^{-3}\,\mathrm{yr}^{-1} f_{\rm PBH}^{53/37}\left(\frac{t}{t_0}\right)^{-34/37} \nonumber\\
    &\quad\times\left(\frac{M}{M_\odot}\right)^{-32/37}S\,\eta^{-34/37}\,\psi(m_1)\psi(m_2)\,,
\end{align}
where $M=m_1+m_2$ is the total mass and $\eta=m_1m_2/M^2$ is the symmetric mass ratio. For a monochromatic mass function with $\psi(m')=\delta(m'-m)$, the present day merger rate reduces to
\begin{equation}
    R_0\simeq 3.14\times10^6\,\mathrm{Gpc}^{-3}\,\mathrm{yr}^{-1}\,S\left(\frac{m}{M_\odot}\right)^{-32/37}f_{\rm PBH}^{53/37}\,,
    \label{eq:mergerrate}
\end{equation}
where the suppression factor is approximated as $S\simeq0.24\left(1+2.3\,\sigma_M^2/f_{\rm PBH}^2\right)^{-21/74}$, with $\sigma_M\simeq0.005$. The equal mass relation $\mathcal{M}=2^{-1/5}m$ provides a direct mapping between the observed chirp mass and the PBH component mass, allowing both the merger rate and detector sensitivity to be expressed in terms of $m$. 
The expected number of detected events during an observing time $T$ is
\begin{equation}
    \mu = \int_0^T {\rm d}t \int {\rm d}V_{\rm eff}\, R(t)\,,
\end{equation}
where ${\rm d}V_{\rm eff}$ denotes the effective comoving volume element ${\rm d}V_c$, weighted by the detection efficiency $p_{\rm eff}$. Since the observing time is negligible compared to cosmological timescales, the merger rate can be approximated as constant, $R(t) \simeq R_0$, yielding~\cite{LVK:2022ydq}
\begin{equation}
    \label{eq:VT}
    \mu \simeq R_0 T \int {\rm d}z\,\frac{\mathrm{d}V_c}{{\rm d}z}\,p_{\rm eff} \equiv R_0\,\langle VT\rangle\,,
\end{equation}
where $\langle VT\rangle$ denotes the effective spacetime volume. This approximation holds for subsolar-mass searches, where the accessible redshift range remains small, but breaks down for higher masses or next-generation detectors probing cosmological distances. Assuming Poisson statistics, the probability of observing at least one merger event under the PBH hypothesis is
\begin{equation}
    \label{eq:probability}
    \mathcal{P}(m) = 1 - e^{-\mu}\,.
\end{equation}
This quantity should not be interpreted as the posterior probability that a given observed event is of primordial origin. In practice, $\langle VT\rangle$ is estimated via Monte Carlo injections as
\begin{equation}
    \langle VT\rangle = T\,\int {\rm d}V_c(\theta)\,p_{\rm det}(\theta)\,,
\end{equation}
where $p_{\rm det}(\theta)$ is the detection probability for binary parameters $\theta$, whose average over the assumed source population gives $p_{\rm eff}$.

To gain analytical insight into the detector sensitivity, we estimate the scaling of $\langle VT\rangle$. The comoving volume element is
\begin{equation}
    \frac{\mathrm{d}V_c}{\mathrm{d}z} = \frac{4\pi cD_c^2(z)}{H(z)}\,,
\end{equation}
where $D_c(z)$ is the comoving distance. For low-mass binaries, the signal is dominated by the inspiral phase. The matched-filter signal-to-noise ratio (SNR) is~\cite{Cutler:1994ys, Maggiore:2007nq, Sathyaprakash:2009xs}
\begin{equation}
    \label{eq:rho2}
    \rho^2 = 4 \int_{f_{\min}}^{f_{\max}}{\rm d}f\,\frac{|\tilde h(f)|^2}{S_n(f)}\,,
\end{equation}
with waveform amplitude approximately given by
\begin{equation}
    |\tilde h(f)|^2 \propto \frac{m^{10/3}}{D_L^2}\,f^{-7/3}\,.
\end{equation}
In the idealized limit of a frequency-independent noise power spectral density, the SNR scales as
\begin{equation}
    \rho \propto \frac{m^{5/6}}{D_L}\,,
\end{equation}
implying a horizon distance $D_h \propto m^{5/6}$ and an effective spacetime volume $\langle VT\rangle \propto m^{5/2}$. In realistic detectors, this scaling is modified by the frequency dependence of the noise, the finite signal bandwidth, and selection effects from search pipelines. We parametrize the SNR scaling as
\begin{equation}
    \rho \propto m^\alpha, \qquad \alpha < 5/6\,,
\end{equation}
which implies $\langle VT\rangle \propto m^{3\alpha}$.

To characterize the detector sensitivity, we use the publicly available sensitive spacetime volume $\langle VT\rangle$ obtained with the Multi-Band Template Analysis (MBTA) search pipeline for the O4a observing run. Although S251112cm occurred during O4c, we adopt the publicly available O4a MBTA sensitivity as a reference for estimating the detection probability. For equal mass binaries, $\mathcal{M}=2^{-1/5}m$, allowing the sensitivity to be expressed directly in terms of $m$. A power-law fit gives
\begin{equation}
    \langle VT\rangle_{\rm O4a}
    \propto m^{\,2.133\pm0.048},
    \qquad \alpha_{\rm O4a}=0.711\pm0.016\,.
\end{equation}
The normalization of $\langle VT\rangle_{\rm O4a}$ is fixed directly by the released MBTA sensitivity values. As a consistency check, the corresponding O3 MBTA data yield $\langle VT\rangle_{\rm O3}\propto m^{\,2.176\pm0.032}$, or $\alpha_{\rm O3}=0.725\pm0.011$, demonstrating that the two observing runs have similar mass dependence and normalization, consistent with $\langle VT\rangle_{\rm O4a}/\langle VT\rangle_{\rm O3}\simeq1$ reported for MBTA~\cite{LIGOScientific:2026wxz,LIGOScientific:2025hdt}. The O4a subsolar search covers detector-frame component masses down to $0.2\,M_\odot$~\cite{LIGOScientific:2026wxz}. We conservatively restrict the equal mass analysis to $0.20\leq m/M_\odot\leq1$. The O4a fit is used only within this interval, so no low-mass extrapolation of the detector sensitivity is performed. Since S251112cm was observed during O4, we use the O4a MBTA sensitivity throughout the following analysis.\footnote{Throughout our analysis, we adopt the central (best-fit) values of the MBTA sensitivity parameters, without propagating their fitting uncertainties.} Adopting the O3 sensitivity leads to essentially unchanged results.

We now assess the impact of relaxing the monochromatic PBH mass assumption by considering a normalized lognormal (LN) mass fraction distribution,
\begin{equation}
    \psi_{\rm LN}(m)= \frac{1}{\sqrt{2\pi}\,\sigma m}\exp\!\left[-\frac{\ln^2(m/m_0)}{2\sigma^2}\right]\,,
    \label{eq:lognormal}
\end{equation}
where $m_0$ is the characteristic mass and $\sigma$ is the logarithmic width. We assume equal mass pairings, approximating the joint mass distribution by $\psi_{\rm LN}(m_1)\delta(m_2-m_1)$. Under this prescription, the differential merger rate is
\begin{equation}
    \begin{aligned}
    \frac{{\rm d}R_{\rm LN}}{{\rm d}m}={}&3.14\times10^{6}\,{\rm Gpc}^{-3}{\rm yr}^{-1}f_{\rm PBH}^{53/37} S_{\rm LN}\\
    &\times\left(\frac{m}{M_\odot}\right)^{-32/37}\psi_{\rm LN}(m) .
    \end{aligned}
    \label{eq:rate_lognormal}
\end{equation}
For an extended mass function, the binary-suppression factor also depends on the PBH mass distribution. Following Refs.~\cite{Raidal:2018bbj, Aljaf:2025dta}, we adopt
\begin{equation}
    S_{\rm LN}\simeq1.42\left[\frac{e^{\sigma^2}}{\bar N+C}+\frac{\sigma_M^2}{f_{\rm PBH}^2}\right]^{-21/74}e^{-\bar N},
    \label{eq:S_LN}
\end{equation}
with $\sigma_M\simeq0.005$ and
\begin{equation}
    \bar N=2\frac{m}{m_0}e^{\sigma^2/2}\frac{f_{\rm PBH}}{f_{\rm PBH}+\sigma_M}\,,
    \label{eq:Nbar_LN}
\end{equation}
where we used the number weighted moments $\langle m\rangle=m_0e^{-\sigma^2/2}$ and $\langle m^2\rangle/\langle m\rangle^2=e^{\sigma^2}$, obtained from the mass-fraction distribution $\psi_{\rm LN}(m)$. The factor $C$ is
\begin{equation}
    C=\frac{f_{\rm PBH}^2e^{\sigma^2}}{\sigma_M^2}\left[{\cal U}^{-74/21}-1\right]^{-1},
    \label{eq:C_LN}
\end{equation}
where
\begin{equation}
    {\cal U}\equiv\frac{\Gamma(29/37)}{\sqrt{\pi}}\,U\!\left(\frac{21}{74},\frac12,\frac{5f_{\rm PBH}^2}{6\sigma_M^2}\right),
\end{equation}
and $U$ is the confluent hypergeometric function.

The microlensing bounds must likewise be recast for an extended mass distribution. Following Ref.~\cite{Carr:2017jsz}, a monochromatic limit $f_{\rm max}^{\rm mono}(m)$ implies
\begin{equation}
    f_{\rm PBH}\int {\rm d}m\,\frac{\psi_{\rm LN}(m)}{f_{\rm max}^{\rm mono}(m)}\leq 1 ,
    \label{eq:extended_constraint}
\end{equation}
and
\begin{equation}
    f_{\rm PBH,max}^{\rm LN} = \left[\int {\rm d}m\, \frac{\psi_{\rm LN}(m)}{f_{\rm max}^{\rm mono}(m)}\right]^{-1}\,.
    \label{eq:fmax_LN}
\end{equation}
As such, both the merger rate suppression and the maximum allowed PBH abundance are modified relative to the monochromatic case.

Using the O4a MBTA sensitivity, the expected number of detectable equal mass mergers in the interval directly covered by the available $\langle VT\rangle$ data is
\begin{equation}
    \mu_{\rm LN} = \int_{0.20M_\odot}^{1M_\odot} {\rm d}m\,\frac{{\rm d}R_{\rm LN}}{{\rm d}m}\,\langle VT\rangle_{\rm O4a}(m)\,,
    \label{eq:mu_lognormal}
\end{equation}
with
\begin{equation}
    \mathcal{P}_{\rm LN}=1-e^{-\mu_{\rm LN}}\,.
\end{equation}
We evaluate representative widths $\sigma=0.5$ and $1$ and compare them with the monochromatic case. In the narrow-width limit, $\sigma \rightarrow 0$, one recovers $\psi_{\rm LN}(m)\rightarrow \delta(m-m_0)$. The resulting impact of the extended mass distribution on the merger probability will be discussed in detail in the following section.

\section{Results}

\begin{figure}[t]
    \centering
    \includegraphics[width=\columnwidth]{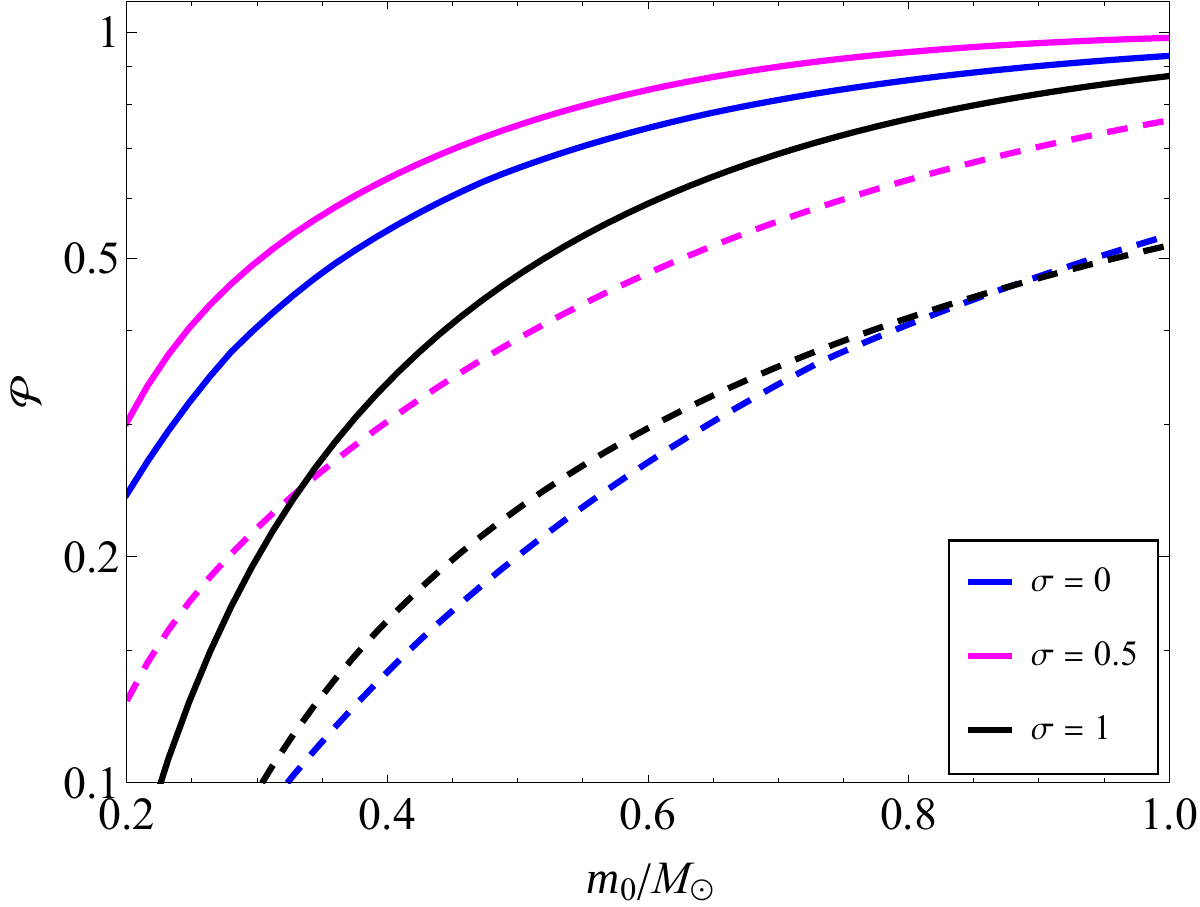}
    \caption{Detection probability $\mathcal{P}$ during O4a as a function of the characteristic PBH mass $m_0$ for a monochromatic mass function, $\sigma=0$ (blue), and lognormal mass functions with $\sigma=0.5$ (magenta) and $\sigma=1$ (black). Solid (dashed) curves correspond to the relaxed (conservative) microlensing constraints.}
    \label{fig:prob_extended_O4a}
\end{figure}

We combine the merger rate prediction with observational constraints on $f_{\rm PBH}$ and the detector sensitivity derived above to compute the probability of observing a PBH merger as a function of mass. We adopt existing microlensing constraints on the PBH abundance over the mass interval relevant to our analysis, $m=0.20$--$1\,M_\odot$~\cite{Mroz:2024mse}, considering both a ``{\it conservative}'' scenario, in which all detected microlensing events are attributed to astrophysical sources, and a more ``{\it relaxed}'' scenario, which is more sensitive to the modeling of the Milky Way disk and halo, for which we adopt the prescription of Ref.~\cite{Cautun:2019eaf, Garcia-Bellido:2024yaz, Hawkins:2025mlo} based on {\it Gaia} DR2.

\begin{figure}[t]
    \centering
    \includegraphics[width=\columnwidth]{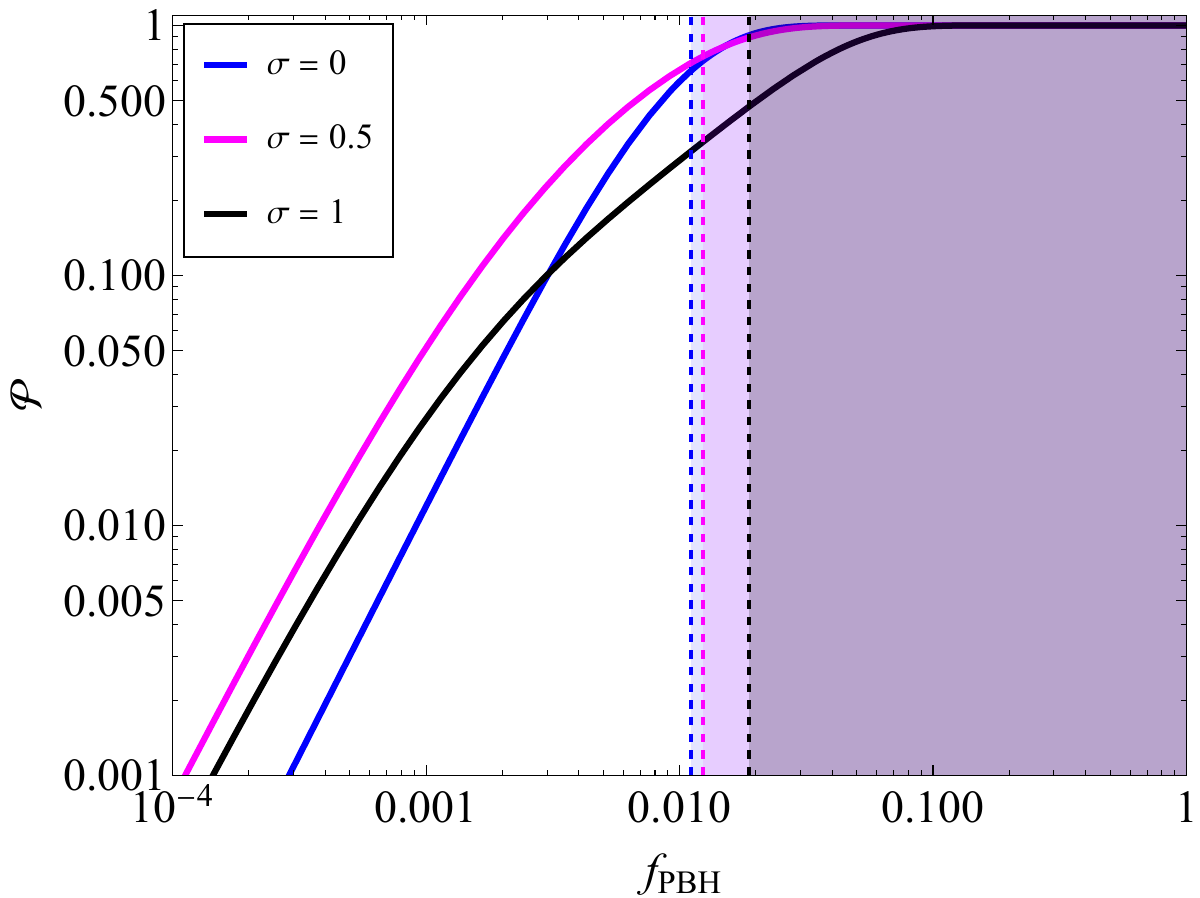}
    \caption{Detection probability $\mathcal{P}$ as a function of the PBH abundance $f_{\rm PBH}$ for the reference characteristic mass $m_0=0.5\,M_\odot$. The blue, magenta, and black curves correspond to $\sigma=0$, $0.5$, and $1$, respectively. The vertical dashed lines indicate the corresponding upper limits on $f_{\rm PBH}$ from the relaxed microlensing constraint, while the lightly shaded regions to their right denote the excluded parameter space.}
    \label{fig:prob_fpbh}
\end{figure}

These constraints set the maximum allowed PBH abundance and directly control the merger probability. Since the O3 and O4a MBTA sensitivities are nearly identical over the mass interval considered here, and S251112cm was observed during O4, we present the results using the O4a sensitivity. Figure~\ref{fig:prob_extended_O4a} shows the resulting probability for the monochromatic case and for lognormal distributions with $\sigma=0.5$ and $1$. For the monochromatic and lognormal scenarios, $\mathcal{P}$ increases with the characteristic mass $m_0$, reflecting the increasing detector sensitivity toward larger component masses. Under the relaxed microlensing constraint, the moderately extended distribution with $\sigma=0.5$ gives the largest probability over most of the mass range and approaches unity toward $m_0\simeq1\,M_\odot$. The broader case $\sigma=1$ is comparatively suppressed, owing to the combined effect of the modified binary-suppression factor and the recast microlensing bound. The conservative constraint systematically lowers the probability, although the same qualitative ordering is retained. Figure~\ref{fig:prob_fpbh} illustrates the dependence on the PBH abundance for the representative value $m_0=0.5\,M_\odot$. In all cases the detection probability rises steeply with $f_{\rm PBH}$ and saturates toward unity at sufficiently large abundance. The relaxed microlensing bounds, however, restrict $f_{\rm PBH}$ to values of order $10^{-2}$, as indicated by the vertical dashed lines; the parameter space to their right is excluded. Importantly, the probability remains sizable at the corresponding allowed upper limits.

\section{Discussion}

Taken together, the two figures show that the observable probability is controlled by the interplay between the PBH abundance, the suppression of the merger rate, the microlensing constraints, and the detector sensitivity, rather than by the width of the mass function alone. In particular, even after imposing the microlensing upper bounds, the probability remains sizable for a subdominant PBH abundance. This shows that the interpretation of S251112cm as a PBH merger remains compatible with current observational constraints.

While remaining sizable throughout the mass range studied, the actual probability, being sensitive to the assumed PBH abundance, depends on astrophysical modeling, which makes the ultimate interpretation of S251112cm as a PBH merger subject to astrophysical uncertainties. Additional microlensing signatures can also arise from axion stars condensed in dark matter minihalos surrounding PBHs~\cite{Yin:2024xov}. Given the current statistical significance of the event, our results should be interpreted as a consistency test of the PBH hypothesis rather than as evidence in its favor.

We emphasize that our analysis assumes equal mass binaries, $m_1=m_2$, both for the monochromatic and extended PBH mass functions considered here. While this assumption provides a simple mapping between the chirp mass, merger rate, and detector sensitivity, an extended mass function generically also allows unequal mass pairings. A complete treatment of such binaries would require integrating over the full two-dimensional mass distribution and could modify both the merger rate and the detection probability. Extended mass functions may also populate both subsolar and higher-mass regimes, potentially linking GW observations to other astrophysical probes, including high-redshift galaxy formation~\cite{Yuan:2023bvh}. The interpretation of JWST observations of high-redshift galaxies also depends on star formation modeling and may involve exotic sources~\cite{Iocco:2024rez}. For a given PBH population, the expected number of detections in future observing runs scales with the sensitive spacetime volume, $\mu=R\langle VT\rangle$. Improved detector sensitivity will increase the accessible merger population, while repeated detections or non-detections will provide increasingly stringent constraints on the PBH abundance and mass distribution.

In addition, some PBH formation scenarios predict a GW background induced by primordial density fluctuations, which can extend into the nanohertz frequency band probed by pulsar timing arrays~\cite{Kohri:2020qqd}. Future GW observations with LVK as well as next-generation detectors such as the Einstein Telescope and Cosmic Explorer will enable one to distinguish between PBHs and alternative compact objects in the subsolar mass range through finite-size effects, such as tidal deformability, which would be large for non-black-hole objects~\cite{Crescimbeni:2024cwh, Crescimbeni:2024qrq}. More broadly, multiple subsolar detections would provide direct information on the underlying PBH mass distribution and abundance, offering a complementary avenue to test the primordial interpretation~\cite{Gouttenoire:2025jxe,Amaral:2026bef}.

Finally, we note that the interpretation of GW events also provides an opportunity to test gravity in the strong-field regime. Deviations from general relativity, such as those predicted in scalar--tensor theories, can modify the waveform and affect both parameter estimation and inferred merger rates. Constraints derived from neutron star--black hole mergers have already been shown to be competitive with traditional probes in certain classes of scalar--tensor models~\cite{Niu:2021nic}. While such effects are subleading for the present analysis, they could become relevant in future high-precision studies of subsolar mass binaries.

\section{Conclusions}

The GW candidate event S251112cm involves at least one compact object of subsolar mass, making its interpretation within standard stellar evolution scenarios challenging. PBHs are natural candidates in this mass range, and we focus on the interval $0.20$--$1\,M_\odot$ directly covered by the O4a subsolar search. We find that the probability of detecting at least one PBH merger during O4a can be substantial and, under the relaxed microlensing constraints, can approach unity for the largest masses considered. By extending the analysis beyond the monochromatic case, we find that a moderately extended lognormal mass function can enhance the detection probability, whereas a broader distribution can instead suppress it. Our results demonstrate that a PBH interpretation of S251112cm remains compatible with current observational constraints.

The possible detection of a subsolar mass compact binary represents a qualitatively new observational window into the origin of black holes, yet care must be taken in the interpretation of these results, which rest on several simplifying assumptions. First, the abundance of PBHs, which we take at the maximum level allowed by current observational constraints, could in fact be lower. Second, although we consider both monochromatic and extended lognormal PBH mass functions, our analysis continues to assume equal mass binaries. For an extended distribution, a more general treatment would allow unequal mass pairings and require an integration over the full two-dimensional mass distribution, which could modify both the merger rate and the detection probability. Mixed binaries involving PBHs and astrophysical black holes could provide an additional possibility. Our results should be regarded as a controlled equal mass realization of the more general PBH scenario.

Both the inferred probability and its extrapolation to future detections depend sensitively on these assumptions. This strengthens the case for future GW observations as a decisive probe of this scenario. The detection of multiple subsolar-mass mergers would provide strong support for a primordial origin, while their absence would further constrain the contribution of PBHs to the dark matter content of the Universe. PBHs remain a viable interpretation of the candidate S251112cm, while GW searches for subsolar binaries offer a direct and complementary probe of early Universe physics and the nature of dark matter. Future observations will be crucial to establish whether such events constitute a genuine population and to robustly assess their origin.

\begin{acknowledgments}
FI and LV acknowledge support by Istituto Nazionale di Fisica Nucleare (INFN) through the Commissione Scientifica Nazionale 4 (CSN4) Iniziativa Specifica ``Theoretical Astroparticle Physics'' (TAsP) and ``Quantum Universe'' (QGSKY).  MRH acknowledges the Tsung-Dao Lee Institute at Shanghai Jiao Tong University for financial support through the Siyuan Postdoctoral Fellowship.
\end{acknowledgments}

\bibliographystyle{apsrev4-1}
\bibliography{references}

\end{document}